\documentclass[twocolumn,a4paper,usenames,accepted=2022-12-21]{quantumarticle} 

\pdfoutput=1
\usepackage[utf8]{inputenc}
\usepackage[english]{babel}
\usepackage[T1]{fontenc}

\usepackage{hyperref}
\usepackage[usenames]{color}
\usepackage{grffile}
\usepackage{graphicx}
\usepackage{amsmath, amstext, amssymb, amsfonts, amsxtra}
\usepackage{textcomp}
\usepackage{xspace}
\DeclareSymbolFontAlphabet{\amsmathbb}{AMSb} 
\usepackage{soul}
\usepackage{xcolor}
\usepackage{epstopdf}
\usepackage{braket}
\usepackage{float}
\usepackage{tikz}
\usepackage{subfigure}
\usepackage{pgfplots}
\usepgfplotslibrary{statistics}
\usepackage{enumitem}
\usepackage[normalem]{ulem}

\setlist{nosep}

\newcommand{\rhop}{\hat{\rho}}

\newcommand{\tr}{{\rm tr}}

\begin{document}

\title{Half-integer vs.\ integer effects in quantum synchronization of spin systems}               

\author{Ryan Tan}
\author{Christoph Bruder}
\affiliation{Department of Physics, University of Basel, Klingelbergstrasse 82, CH-4056 Basel, Switzerland}    
\author{Martin Koppenh{\"o}fer}
\affiliation{Pritzker School of Molecular Engineering, University of Chicago, Chicago, Illinois 60637, USA}

\begin{abstract}
We study the quantum synchronization of a single spin driven by an external semiclassical signal for spin numbers larger than $S=1$, the smallest system to host a quantum self-sustained oscillator. 
The occurrence of interference-based quantum synchronization blockade is found to be qualitatively different for integer vs.\ half-integer spin number $S$.
We explain this phenomenon as the interplay between the external signal and the structure of the limit cycle in the generation of coherence in the system.
Moreover, we show that the same dissipative limit-cycle stabilization mechanism leads to very different levels of quantum synchronization for integer vs.\ half-integer $S$.
However, by choosing an appropriate limit cycle for each spin number, comparable levels of quantum synchronization can be achieved for both integer and half-integer spin systems.
\end{abstract}

\maketitle

\section{Introduction}\label{sec:intro}      
  
From ticking metronomes to blinking fireflies, the phenomenon of synchronization occurs in a wide variety of systems~\cite{pikovsky2003sync}. 
Classically, synchronization has been studied since the 17th century and has found applications in many areas of our daily lives, such as in time-keeping devices and power grids~\cite{strogatz2012sync}. Studies of synchronization in the quantum regime started only recently and have raised a number of fascinating and fundamental questions. These include the definition, existence, and measurement of quantum synchronization, its relation to other measures of quantumness, and the prediction of quantum effects in synchronization that have no classical analogue.  After early pioneering work on the quantum kicked rotator~\cite{Shepelyansky2006}, the synchronization behavior of a variety of quantum systems has been investigated. This includes optomechanical systems~\cite{Ludwig2013}, masers~\cite{davis2016synchronization}, and the quantum van der Pol oscillator \cite{Lee2013,marquardtPRA2017,lorch2017quantum,amitai2017synchronization,amitai2018quantum,es2020synchronization,Wachtler2022}, whose classical counterpart has long been an important model in studies of non-linear dynamics~\cite{strogatz2015nonlinear}.

More recently, synchronization of quantum spins has been studied \cite{Goychuk2006,Zhirov2008,Giorgi2013,Xu2014,Ameri2015,RouletBruderPRL2018a,RouletBruderPRL2018b,KoppenhoferPRA2019}. 
This is in part motivated by the simple structure of spins, viz.\ their finite-dimensional Hilbert space, that enables tackling fundamental questions in quantum synchronization, and in part by potential experimental implementations of quantum synchronization~\cite{PhysRevX.5.021026,neeley2009emulation,koppenhofer2020quantum,ghoshprl2020,Krithika2021}. 
While low-spin systems provide a convenient platform to study the basic principles of quantum synchronization \cite{RouletBruderPRL2018a,RouletBruderPRL2018b,KoppenhoferPRA2019}, the question arises how quantum effects in synchronization change if one varies the spin number between the smallest possible value of $S=1$ \cite{RouletBruderPRL2018a} and large quasi-classical values. 
Moreover, it is an open issue whether there are differences in the synchronization behavior of integer vs.\ half-integer spin systems.

In this paper, we address these open issues by analyzing different spin-$S$ limit-cycle oscillators subject to a resonant semiclassical signal. 
For a given limit-cycle stabilization mechanism, we find qualitative differences in the synchronization behavior of half-integer vs.\ integer spins. 
While this is reminiscent of the famous spin-statistics theorem, we show that these differences can actually be traced back to the presence (absence) of an eigenstate $\ket{S,0}$ of the $\hat{S}_z$ operator with eigenvalue zero for integer (half-integer) spin number $S$. 
The presence or absence of the state $\ket{S,0}$ is also at the heart of other integer vs.\ half-integer effects in quantum optics \cite{Agarwal1989,Agarwal1990}, phase transitions in the Lipkin-Meshkov-Glick model \cite{Unanyan2003,Vidal2004}, and quantum metrology \cite{Molmer1999,Leibfried2004,Leibfried2005,Koppenhoefer2022}.

An important quantum effect is the phenomenon of interference-based quantum synchronization blockade, which is the absence of synchronization due to destructive interference of coherences, even though a weak harmonic signal is applied to the limit-cycle oscillator \cite{KoppenhoferPRA2019}. 
It occurs at specific values of the relative strength of the dissipation rates, and we find that the cases of integer and half-integer spins differ in the number of synchronization blockades that can be observed. 
This is due to the interplay between the way a semiclassical signal builds up coherence in the spin system and the structure of the limit cycle for different spin numbers.

Our results show that there is indeed a difference between integer and half-integer spins if limit cycle and signal are kept fixed while the spin number $S$ is changed.
However, we also find that the situation is entirely different if one is allowed to change the limit cycle for each $S$. 
In this case, a half-integer vs.\ integer effect appears to be absent in the maximum amount of quantum synchronization that can be achieved in a spin-$S$ system.

This paper is organized as follows. We first present the model and framework of quantum synchronization in Sec.~\ref{sec:Singlespin}, highlighting the difference in limit-cycle stabilization between integer spins and half-integer spins. 
Subsequently, in Sec.~\ref{sec:ELC}, we present results on the synchronization of a spin-$\frac{3}{2}$ and spin-$2$ system
and show how the anti-symmetrical generation of coherences leads to synchronization blockade. In Sec.~\ref{sec:epsilon2} and Sec.~\ref{sec:nELC}, we show respectively how the choice of a different signal strength and a different limit cycle affect synchronization. 
In Sec.~\ref{sec:Compare}, we discuss our numerical results for the synchronization measure for different spin values ($S=1$ to $S=3$) and different (types of) limit cycles and compare them to an analytical upper bound that is derived in App.~\ref{app:C}.
Finally, we summarize and draw our conclusions in Sec.~\ref{sec:conclusions}.

\section{Model}\label{sec:Singlespin}

In classical physics, a limit-cycle oscillator is a system that is excited into self-sustained periodic motion with a free phase by an internal source of energy~\cite{pikovsky2003sync}. 
When different limit-cycle oscillators are weakly coupled, they may adjust their oscillations to a common frequency, which is called \emph{mutual synchronization}. 
Alternatively, a single limit-cycle oscillator may adjust its oscillation to a weak external signal, which is called \emph{entrainment}. 
Different ways to generalize these concepts to the quantum realm have been proposed, e.g., using information-theoretic measures to quantify correlations between mutually synchronized limit-cycle oscillators \cite{Mari2013,Lee2014,Ameri2015,Galve2017,Jaseem2020} as well as coherence in an entrained oscillator \cite{Jaseem2020}, and measures of the localization of (relative) phases \cite{Lee2013,Hush2015,Ludwig2013,Weiss2016}.
We consider a quantum limit-cycle oscillator implemented in a spin-$S$ system and use a framework that generalizes the definition of a classical limit-cycle oscillator to the quantum realm based on its phase-space dynamics \cite{RouletBruderPRL2018a,KoppenhoferPRA2019}.

\begin{figure}
    \centering
    \includegraphics[scale=0.3]{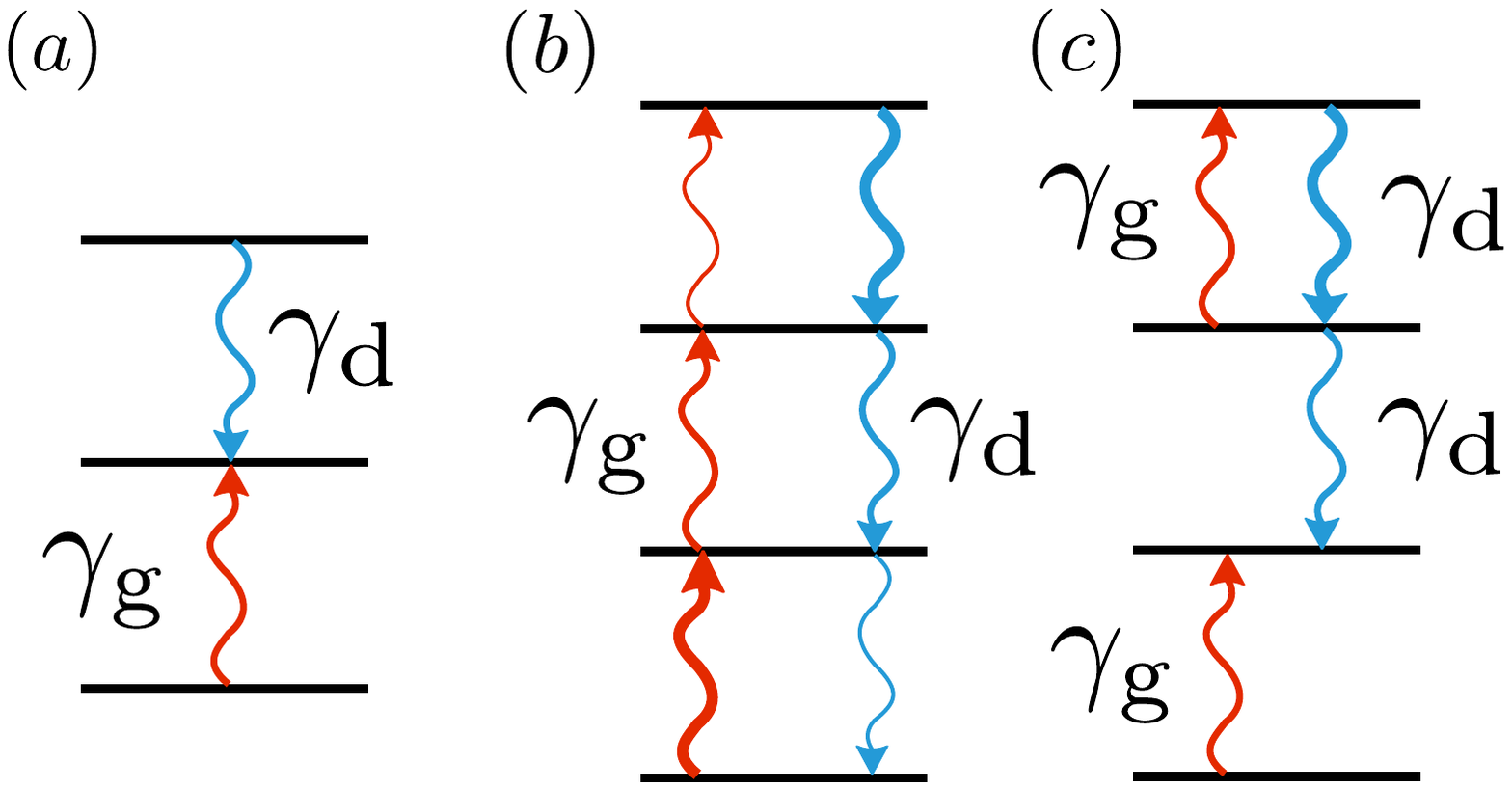}
	\caption{
		Schematic of a spin-$1$ limit-cycle oscillator stabilized to the equatorial state $\ket{S=1,m=0}$ by the SymLC stabilization scheme (a) and a spin-$\frac{3}{2}$ limit-cycle oscillator with two different stabilization schemes (b), (c). 
		Wavy lines represent dissipative processes: red lines indicate gain processes while blue lines represent loss processes; the interplay of the two kinds of processes defines the limit-cycle state. 
		The thickness of the lines corresponds to the relative transition rates of the individual oscillators. 
		(b) SymLC stabilization scheme as defined by Eq.~\eqref{eq:me1spin}, with the steady state dependent on the values of $\gamma_\mathrm{g}$ and $\gamma_\mathrm{d}$. (c) Limit cycle stabilized by the set of modified jump operators $\hat{S}_{\pm}(\hat{S}_z+1/2)$ (AsymLC scheme) resulting in the steady state $\ket{3/2,-1/2}$, independent of the values of $\gamma_\mathrm{g}$ and $\gamma_\mathrm{d}$.
	}
	\label{fig:spin32Lvls}
\end{figure}

In this framework, a 
quantum limit-cycle oscillator is a quantum system which oscillates with a certain amplitude at its natural frequency $\omega_0$. 
The amplitude is stabilized by dissipative gain and loss processes such that the phase of oscillation is completely free and can be adjusted by a weak external signal with a (potentially different) frequency $\omega$. 
In a spin system, the natural frequency of oscillation of the limit-cycle oscillator is set by the level splitting $\omega_0$.
We consider an external semiclassical signal of amplitude $\varepsilon$ and frequency $\omega$, such that the Hamiltonian of the system in a frame rotating at the signal frequency is given by
\begin{align}
\hat{H} = \Delta  \hat{S}_z + \varepsilon \hat{S}_y~.
\label{eq:Ham1spin}
\end{align}
Here, $\Delta = \omega_0 - \omega$ is the detuning between the external signal and the natural frequency of oscillation. In this paper, we focus on the resonant limit $\Delta=0$ for all numerical simulations. $\hat{S}_j$, $j \in \{x,y,z\}$, are the spin operators of a spin-$S$ system, which fulfill the commutation relations $ [\hat{S}_i,\hat{S}_j] = i \epsilon_{ijk}\hat{S}_k$.
A basis of the Hilbert space is given by the states $\ket{S,m}$, $m \in \{-S, \dots, S\}$, which are the joint eigenstates of $\mathbf{\hat{S}}^2 = \hat{S}_x^2 + \hat{S}_y^2 + \hat{S}_z^2$ and $\hat{S}_z$.

To visualize the state of the limit-cycle oscillator and to quantify synchronization, it is instructive to use a phase-space representation.
Following Ref.~\cite{RouletBruderPRL2018a}, we use the Husimi $Q$ function, which is defined for a spin-$S$ system as
\begin{align}
Q(\theta,\phi) = \frac{2S+1}{4\pi}\bra{\theta,\phi}\rhop\ket{\theta,\phi}~.
\label{eq:Qfunc}
\end{align}
Apart from a scale factor, Eq.~\eqref{eq:Qfunc} is the expectation value of the density matrix $\rhop$ with respect to spin-coherent states 
$\ket{\theta,\phi} = \exp(-i\phi\hat{S}_z)\exp(-i\theta\hat{S}_y)\ket{S,S}$ \cite{radcliffe1971}. 
The angles $\theta$ and $\phi$ parameterize the amplitude and the phase of the state respectively. 
A measure of quantum synchronization can thus be obtained from the marginal distribution of the phase, 
\begin{align}
S(\phi) = \int^\pi_0 d\theta \sin\theta\: Q(\theta,\phi) -\frac{1}{2\pi}~.
\label{eq:Sq1}
\end{align}
The last term subtracts the value of a uniform phase distribution, such that $S(\phi)$ is zero  everywhere in the absence of synchronization, while a positive value $S(\phi)>0$ indicates phase accumulation at $\phi$. 
Explicit expressions for $Q(\theta,\phi)$ and $S(\phi)$ for arbitrary spin $S$ are given in App.~\ref{app:A} and below.
Being a probability distribution for the phase $\phi$, $S(\phi)$ offers a straightforward interpretation and can be used to identify switching between multiple synchronized states at different phases (which otherwise requires simultaneous use of multiple synchronization measures \cite{Weiss2016}). 
If needed, Eq.~\eqref{eq:Sq1} can be further reduced to a single-number measure of quantum synchronization by considering its maximum value $\max_\phi S(\phi)$.
An upper bound on $\max_\phi S(\phi)$ is derived in App.~\ref{app:C}.
Note that Eqs.~\eqref{eq:Qfunc} and~\eqref{eq:Sq1} can be generalized to multipartite systems, which allows one to quantify relative phases in the context of mutual synchronization.
Scenarios with dissimilar subsystems, as discussed in Ref.~\cite{Jaseem2020}, can also be addressed if a function $Q$ can be defined for each subsystem.

The amplitude of oscillation is stabilized by incoherent gain and loss processes that can be modeled by a Lindblad master equation
\begin{align}
\dot{\rhop} = -i [\hat{H},\rhop] + \gamma_\mathrm{g}\mathcal{D}[\hat{O}_\mathrm{g}]\rhop + \gamma_\mathrm{d}\mathcal{D}[\hat{O}_\mathrm{d}]\rhop\:, \label{eq:me1spin} 
\end{align}
where the dissipators are defined as $\mathcal{D}[\hat{O}]\rhop = \hat{O}\rho\hat{O}^\dag - \frac{1}{2}\{\hat{O}^\dag\hat{O},\rhop\}$. 
The jump operators $\hat{O}_\mathrm{g}$ and $\hat{O}_\mathrm{d}$ represent the incoherent gain and loss processes at the respective rates $\gamma_\mathrm{g}$ and $\gamma_\mathrm{d}$. 
The structure of the limit cycle (which is the steady state of Eq.~\eqref{eq:me1spin} in the absence of a signal, $\varepsilon = 0$) depends on the specific choice of the jump operators $\hat{O}_\mathrm{g}$ and $\hat{O}_\mathrm{d}$. 
They must be invariant up to a phase factor under rotations about the quantization axis (in our case, the $z$-axis) to avoid any phase preference of the limit cycle \cite{KoppenhoferPRA2019}.
The extremal states $\ket{S,\pm S}$ have zero amplitude, therefore, a valid limit cycle should always include a non-extremal state $\ket{S,m}$ with $|m| < S$.
This precludes spin-$1/2$ systems to host a limit-cycle oscillation in this framework \cite{RouletBruderPRL2018a}. In addition, spin-$1/2$ systems cannot exhibit the interference-based quantum synchronization blockade effect we are interested in. 
Note that one can still define effective descriptions of quantum limit-cycle oscillators in terms of two-level systems by excluding certain spin states from the dynamics \cite{Lee2014}, by focusing only on the phase dynamics \cite{Goychuk2006}, or by considering a different definition of a quantum limit-cycle oscillator \cite{Galve2017,buca2021algebraic}.

The simplest set of jump operators which fulfills all the requirements mentioned above for a spin-$1$ system is given by $\hat{O}_\mathrm{g} = \hat{S}_+\hat{S}_z$ and $\hat{O}_\mathrm{d} = \hat{S}_-\hat{S}_z$, where $\hat{S}_\pm = \hat{S}_x \pm i \hat{S}_y$ are the spin ladder operators. 
The jump operator $\hat{O}_\mathrm{d}$ ($\hat{O}_\mathrm{g}$)
provides no transition rate downwards (upwards) from the level $\ket{1,0}$, i.e., population is transferred towards the $\ket{1,0}$ level and subsequently trapped there, see Fig.~\ref{fig:spin32Lvls}(a).
As a result, the limit cycle is independent of the specific values of $\gamma_\mathrm{g}$ and $\gamma_\mathrm{d}$. 
More generally, for any \emph{integer} spin $S$, this particular set of jump operators stabilizes the limit-cycle state $\ket{S,0}$, which is localized at the equator of a spherical projection of the Husimi $Q$ function.

In contrast, the same set of dissipators stabilizes a \emph{half-integer} spin very differently. The transition rates between the spin levels due to the jump operator $\hat{S}_- \hat{S}_z$ ($\hat{S}_+ \hat{S}_z$) decrease down (up) the ladder of spin states, but every level has nonzero transition rates to neighboring states, as shown in Fig.~\ref{fig:spin32Lvls}(b). 
As a consequence, population is not trapped in a particular spin state $\ket{S,m}$ and the steady state is generally dependent on the dissipation rates.

Note that there is a symmetry between the gain and loss processes: for each gain process up the ladder of states, say, from $\ket{S,m}$ to $\ket{S,m+1}$, there is a corresponding loss process down the ladder from $\ket{S,-m}$ to $\ket{S,-m-1}$, scaled by their respective rates $\gamma_\mathrm{g}$ and $\gamma_\mathrm{d}$. This symmetry is present for both integer and half-integer spins [Fig.~\ref{fig:spin32Lvls}(a) and (b)].
Therefore, we refer to the set of dissipators $\hat{O}_\mathrm{g} = \hat{S}_+ \hat{S}_z$, $\hat{O}_\mathrm{d} = \hat{S}_- \hat{S}_z$ as the \textit{gain-loss-symmetric limit-cycle} (SymLC) stabilization scheme in the following.

\begin{figure}[t]
	\centering
    \includegraphics[width=\columnwidth]{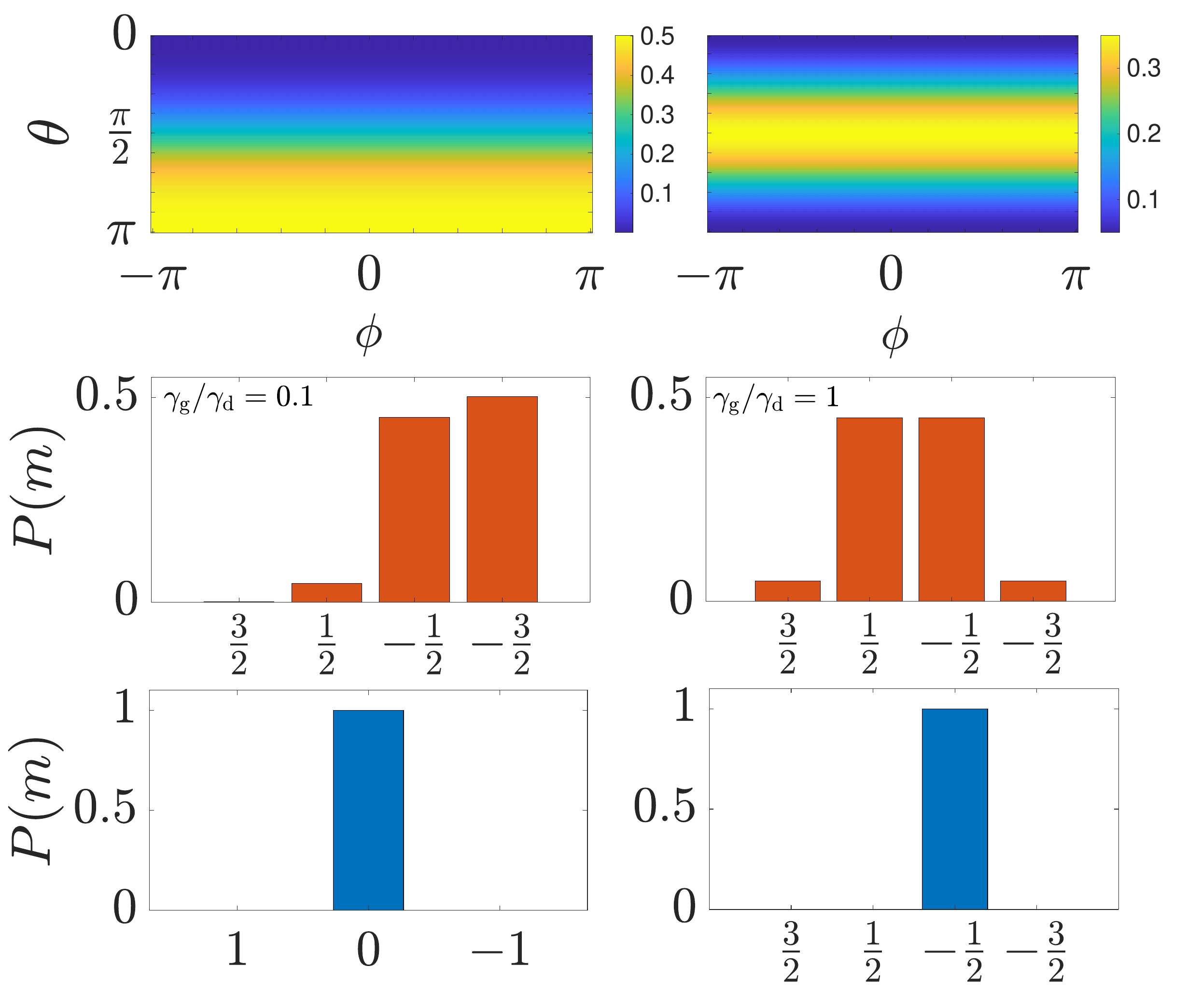}
	\caption{
	Husimi $Q$ function (top) and population $P(m)$ of the spin states $\ket{S,m}$ (bottom) for the limit-cycle state for spins $S=3/2$ and $S=1$ (last row, left). 
	The $m$-th level corresponds to the states $m=\{3/2,1/2,-1/2,-3/2\}$ for $S=3/2$ and $m=\{1,0,-1\}$ for $S=1$ respectively. 
	For $S=3/2$, the uniform phase distribution (first row)	indicates a valid limit cycle for both values of dissipation rates, $\gamma_\mathrm{g}/\gamma_\mathrm{d} = 0.1$ (left) and $\gamma_\mathrm{g}/\gamma_\mathrm{d} = 1$ (right). 
	The second row (red) clearly shows the dependence of the populations $P(m)$ on the dissipation rates in a half-integer spin system with the SymLC stabilization scheme sketched in Fig.~\ref{fig:spin32Lvls}(b).
	In contrast, in the third row (blue) that corresponds to the SymLC stabilization schemes for a spin $S=1$ [sketched in Fig.~\ref{fig:spin32Lvls}(a)] and the AsymLC stabilization scheme for a spin $S=3/2$ [sketched in Fig.~\ref{fig:spin32Lvls}(c)],
	the limit-cycle states are independent of the dissipation rates.} 
	\label{fig:Fig1}
\end{figure}

As an illustration, the limit cycle stabilized by these operators in a spin-$3/2$ system has the form
\begin{align}
\hat{\rho}_\mathrm{ss} = \mathcal{N}\text{diag} \left(\frac{\gamma_\mathrm{g}^3}{\gamma_\mathrm{d}^3},9\frac{\gamma_\mathrm{g}^2}{\gamma_\mathrm{d}^2},9\frac{\gamma_\mathrm{g}}{\gamma_\mathrm{d}},1 \right),
\label{eqn:LC32}
\end{align}
where $\mathcal{N}$ is a normalization constant such that $\tr(\hat{\rho}_\mathrm{ss})=1$.  
For balanced dissipation rates $\gamma_\mathrm{g}/\gamma_\mathrm{d} = 1$, the statistical weights of the $\ket{3/2,\pm 1/2}$ states are the greatest and the Husimi $Q$ function is centered about the equator, as shown in Fig.~\ref{fig:Fig1}.
For imbalanced dissipation rates $\gamma_\mathrm{g}/\gamma_\mathrm{d}=0.1$, however, the statistical weights of the states $\ket{3/2,-1/2}$ and $\ket{3/2,-3/2}$ are the greatest and the Husimi $Q$ function is localized in the vicinity of the south pole. The gain-loss symmetry is evident when one applies the exchange transformation $\gamma_\mathrm{g}\leftrightarrow\gamma_\mathrm{d}$, resulting in $\rho_{m,m} = \rho_{-m,-m}$.

In the next section, we analyze the different synchronization behavior of the SymLC scheme for integer and half-integer spins. 
In Section~\ref{sec:nELC}, we show that by slightly modifying the dissipators and breaking the symmetry between gain and loss processes, we can stabilize the limit cycle to a non-equatorial state which can lead to qualitatively different quantum synchronization behavior.

\section{Gain-loss-symmetric limit-cycle stabilization}\label{sec:ELC}

In this section, we compare the synchronization behavior of the SymLC stabilization scheme for different spin numbers $S$. 
For a spin-$1$ system, it has been shown that the SymLC is unable to synchronize to a semiclassical drive if its dissipation rates are balanced, $\gamma_\mathrm{g} /\gamma_\mathrm{d} = 1$, due to destructive interference of the coherences built up by the signal \cite{KoppenhoferPRA2019}. 
This so-called synchronization blockade can be lifted by choosing imbalanced dissipation rates, $\gamma_\mathrm{g}/\gamma_\mathrm{d} \neq 1$. 
We now show that higher spins $S> 1$ exhibit an even richer synchronization blockade behavior. 
To gain an initial understanding of how synchronization changes with the next larger spin values, we fix the external signal strength relative to the minimum of the dissipation rates, 
\begin{align}
\varepsilon_1=\eta\min (\gamma_\mathrm{g},\gamma_\mathrm{d}).
\label{eq:epsilon1}
\end{align}
Here, $\eta$ is a small parameter chosen such that the external signal remains a perturbative effect on the limit cycle.

\begin{figure}[t]
	\centering
	\includegraphics[width=\columnwidth]{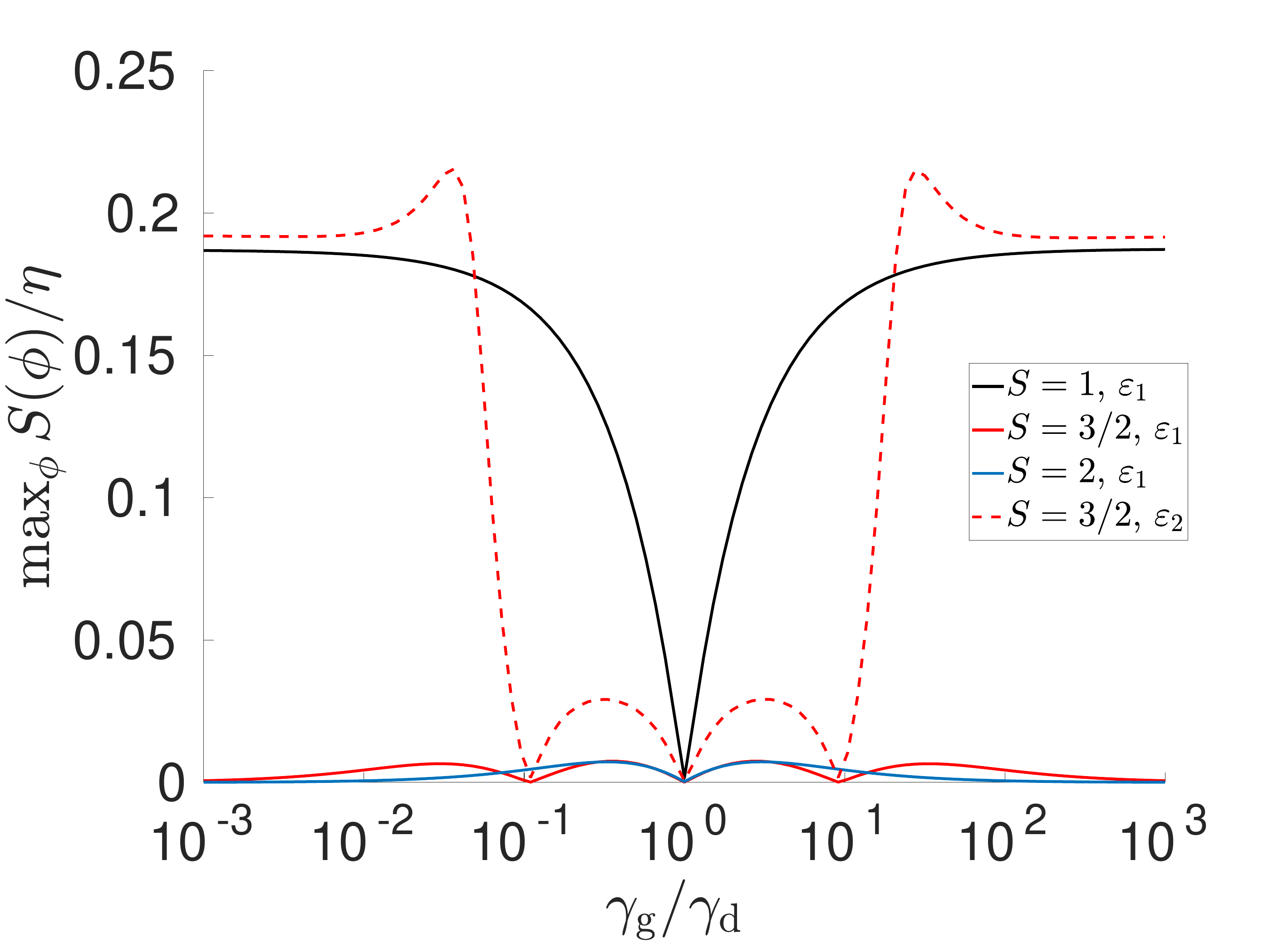}
    	\caption{Maximum of phase distribution $S(\phi)/\eta$ vs.\ ratio of dissipation rates $\gamma_\mathrm{g}/\gamma_\mathrm{d}$, for spin values $S=1$ (black), $3/2$ (red), and $2$ (blue), under the SymLC stabilization scheme and a semiclassical signal. The synchronization blockade persists at balanced rates $\gamma_\mathrm{g}/\gamma_\mathrm{d} = 1$ for all spin values, with an additional blockade for $S=3/2$ appearing at intermediate ratios $\gamma_{g(d)}/\gamma_{d(g)} \sim 10^{-1}$. By defining the  signal strength $\varepsilon_2$ as described in Sec.~\ref{sec:epsilon2}, a larger signature of synchronization is achieved in general. In particular, for $S=3/2$, the $\varepsilon_2$ drive lifts the blockade at large dissipation ratios $\gamma_{g(d)} \ll \gamma_{d(g)}$ (dashed line). The other parameters are $\eta = 0.01$ and $\Delta = 0$.
	}
	\label{fig:ELC_epsilon1}
\end{figure}

In Fig.~\ref{fig:ELC_epsilon1}, we have plotted $\max_\phi S(\phi)$ over a large range of ratios of dissipation rates for the different spin values. For all studied spin numbers ($S=1,3/2,2)$, we observe that $\max_\phi S(\phi)$ vanishes at $\gamma_\mathrm{g}=\gamma_\mathrm{d}$ corresponding to the synchronization blockade known from $S=1$. However, we see that unlike in the $S=1$ case, choosing imbalanced dissipation rates never completely lifts the synchronization blockade for both $S=3/2$ and $S=2$, and the synchronization measure remains suppressed for largely imbalanced rates. Furthermore, we observe an additional suppression of the synchronization measure at an intermediate ratio of dissipation rates for $S=3/2$. These phenomena can be understood by studying how coherences are generated as a result of the semiclassical drive.

\subsection{Synchronization measure for arbitrary spin number $S$}

To understand the connection between the coherences and the measure of quantum synchronization, we express the marginal phase distribution $S(\phi)$ for arbitrary spin $S$ 
as follows
\begin{align}\label{eq:SphiGeneral}
S(\phi) = \sum_{m,m'=-S}^S e^{-i(m-m')\phi}d^{S}_{m,m'}\rho_{m',m}.
\end{align}
The coefficients $d^{S}_{m,m'}$ can be expressed in terms of elements of the Wigner D-matrix, with the amplitude degree of freedom $\theta$ integrated out (see App.~\ref{app:A} for details and explicit expressions). 
There are $2S$ oscillatory terms in Eq.~\eqref{eq:SphiGeneral}, each corresponding to a particular wavenumber $k = m - m'$. 
The amplitude of each of these terms depends on the prefactors $d^{S}_{m,m'}$ and the coherences $\rho_{m',m} = \bra{S,m'} \hat{\rho} \ket{S,m}$ between all spin eigenstates $\ket{S,m'}$ and $\ket{S,m}$ with $k = m - m'$. 
In addition, the prefactors fulfill the relation $d^{S}_{m,m'} = d^{S}_{-m',-m}$, implying that the coherences $\rho_{m',m}$ and $\rho_{-m,-m'}$ contribute equally to $S(\phi)$. 
In the special case when all those pairs of coherences interfere destructively, $\rho_{m',m}= -\rho_{-m,-m'}$, the amplitude of the term  $e^{-i k \phi}$ vanishes and the synchronization measure is suppressed even though each individual coherence is nonzero. 
We show next that the points where the synchronization measure is suppressed in Fig.~\ref{fig:ELC_epsilon1} are precisely caused by such an interference-based quantum synchronization blockade.

\subsection{Coherence generation in the SymLC stabilization scheme}
\label{sec:CoherenceGeneration}

Projecting Eq.~\eqref{eq:me1spin} onto the spin eigenstates $\ket{S,m}$, the steady-state coherence between adjacent spin states, to leading order $\varepsilon$, is
\begin{widetext}
\begin{align}
	\rho_{n,n - 1} &= \frac{\varepsilon A^-_n }{2 i \Delta + \gamma_\mathrm{d} \left[ (n-1)^2 (A^-_{n-1})^2 + n^2 (A^-_{n})^2 \right] + \gamma_\mathrm{g} \left[ (n-1)^2 (A^+_{n-1})^2 + n^2 (A^+_{n})^2 \right]} \times \left[ \rho_{n,n} - \rho_{n-1,n-1} \right]~,
	\label{eqn:CoherenceIntegerSpin}
\end{align} 
\end{widetext}
where $A^\pm_m = \sqrt{S(S+1) - m (m \pm 1)}$ denotes the matrix elements of the spin raising and lowering operators, $\hat{S}_\pm \ket{S,m} = A^\pm_{m} \ket{S, m \pm 1}$. 
A derivation of this expression is given in App.~\ref{app:B}. 
We used the result that, to leading order in the signal strength, only the first off-diagonal coherences are built up, $\rho_{n,n\pm 1} = \mathcal{O}(\varepsilon)$ but $\rho_{n,n \pm k} = \mathcal{O}(\varepsilon^2)$ for $k \geq 2$. Additionally, we used the identity $A^+_{m-1} = A^-_m$ to simplify the expression. 

For a spin-$1$ SymLC, this expression reproduces the known results \cite{KoppenhoferPRA2019}
\begin{align}
	\rho_{1,0} &= - \varepsilon \frac{1}{\sqrt{2}(\gamma_\mathrm{d} +  i \Delta)} ~, \\
	\rho_{0,-1} &= + \varepsilon \frac{1}{\sqrt{2}(\gamma_\mathrm{g} +  i \Delta)} ~.
\end{align}
These coherences cancel out exactly for $\gamma_\mathrm{g} = \gamma_\mathrm{d}$. 
In the limit $\gamma_\mathrm{g}/\gamma_\mathrm{d} \ll 1$ and for a resonant semiclassical signal with the signal strength defined by Eq.~\eqref{eq:epsilon1}, the coherence $\rho_{1,0} \propto \eta \gamma_\mathrm{g}/\gamma_\mathrm{d} \lll 1$ is strongly suppressed whereas the other coherence tends to the constant value $\rho_{0,-1} = \eta /\sqrt{2}$. 
The perfect destructive interference at $\gamma_\mathrm{g} = \gamma_\mathrm{d}$ is thus lifted and the synchronization measure increases. 
The same effect occurs in the opposite limit $\gamma_\mathrm{g}/\gamma_\mathrm{d} \gg 1$, where $\rho_{0,-1} \propto \eta \gamma_\mathrm{d}/\gamma_\mathrm{g} \lll 1$ is suppressed but $\rho_{1,0} = - \eta/ \sqrt{2}$.

We now consider higher integer spins, $S \geq 2$. 
Equation~\eqref{eqn:CoherenceIntegerSpin} shows that they must exhibit the same synchronization blockade for balanced dissipation rates $\gamma_\mathrm{g} = \gamma_\mathrm{d}$: 
The first factor in Eq.~\eqref{eqn:CoherenceIntegerSpin} is invariant under the simultaneous transformation 
\begin{align}
	n = S - m &\leftrightarrow n = -S + m + 1 ~, \nonumber\\
	\gamma_\mathrm{g} &\leftrightarrow \gamma_\mathrm{d} ~, \label{eqn:SymmetryTrafo}
\end{align}
because the matrix elements $A^\pm_m$ satisfy the symmetry relation $A^+_{S-m} = A^-_{-S+m}$.  
This transformation ``inverts'' the ladder of spin states $\ket{S,m}$ about the state $\ket{S,0}$, i.e, it exchanges the populations $\rho_{1,1} \leftrightarrow \rho_{-1,-1}$, $\rho_{2,2} \leftrightarrow \rho_{-2,-2}$, etc., as well as the coherences $\rho_{1,0} \leftrightarrow \rho_{0,-1}$, $\rho_{2,1} \leftrightarrow \rho_{-1,-2}$, etc. 
Since the limit cycle is given by the state $\ket{S,0}$, we have $\rho_{n,n} = \delta_{n,0}$ and the second factor changes sign under this transformation.
Therefore, every pair of ``opposite'' coherences cancels. 
More generally, any limit cycle that is symmetric under the transformation~\eqref{eqn:SymmetryTrafo} will show interference-based quantum synchronization blockade for balanced dissipation rates since
\begin{align}
	&\frac{\rho_{S-m,S-m-1}}{\rho_{-S+m+1,-S+m}\vert_{\gamma_\mathrm{g} \leftrightarrow \gamma_\mathrm{d}}} \nonumber \\
	&= - \frac{\rho_{S-m,S-m} - \rho_{S-m-1,S-m-1}}{[\rho_{-S+m,-S+m} - \rho_{-S+m+1,-S+m+1}]_{\gamma_\mathrm{g} \leftrightarrow \gamma_\mathrm{d}}}\nonumber \\
	&= -1 ~.
	\label{eqn:SymmetryIntegerSpin}
\end{align}
The left-hand side is the ratio between two coherences which are equally many levels below and above the extremal spin states $\ket{S,S}$ and $\ket{S,-S}$, respectively. 
The right-hand side relates this ratio to the difference between the populations of the spin levels connected by the two coherences, i.e., it depends on the properties of the limit-cycle state.

Equation~\eqref{eqn:CoherenceIntegerSpin} also reveals another difference between the case of $S=1$ and higher integer spins. 
For any $S \geq 2$, all coherences $\rho_{n,m}$ will in general depend on both $\gamma_\mathrm{d}$ and $\gamma_\mathrm{g}$ because both jump operators $\hat{O}_\mathrm{g}$ and $\hat{O}_\mathrm{d}$ mediate transitions away from the associated spin levels $\ket{S,n}$ and $\ket{S,m}$.
For instance, for $S=2$, we have
\begin{align}
\label{eq:spin2Coherence}
\rho_{1,0} &= - \varepsilon \frac{\sqrt{3/2}}{3\gamma_\mathrm{d}+2\gamma_\mathrm{g}+i\Delta}~, \\    
\rho_{0,-1} &= + \varepsilon \frac{\sqrt{3/2}}{3\gamma_\mathrm{g}+2\gamma_\mathrm{d}+i\Delta}~.
\end{align}
The presence of both decay rates in the denominator implies that, for imbalanced dissipation rates and a signal strength $\varepsilon_1$ defined by Eq.~\eqref{eq:epsilon1}, both coherences will scale proportionally to $ \min(\gamma_\mathrm{g},\gamma_\mathrm{d})/\max(\gamma_\mathrm{g},\gamma_\mathrm{d})$. 
As a result, at largely imbalanced dissipation rates, synchronization will be suppressed, in contrast to the $S=1$ case where it converged to a constant value determined by one of the two coherences.

Note that for integer spin, a semiclassical signal, and the SymLC stabilization scheme, there is only one synchronization blockade at balanced dissipation rates. 
For $\gamma_\mathrm{g} \neq \gamma_\mathrm{d}$, the first term in Eq.~\eqref{eqn:CoherenceIntegerSpin} loses its inversion symmetry but the limit cycle remains fixed and thus keeps its inversion symmetry. 
Therefore, no further cancellations of coherences are expected.

This changes if we now move on to the case of half-integer spins, $S \geq 3/2$. 
Equation~\eqref{eqn:CoherenceIntegerSpin} is valid for half-integer spins, too, and, for balanced dissipation rates, the SymLC stabilization scheme generates a limit cycle which is symmetric under the transformation~\eqref{eqn:SymmetryTrafo}.
Therefore the coherences form pairs of opposite sign which interfere destructively, similar to the case of integer spin. 
Note that the only unpaired coherence $\rho_{1/2,-1/2} = 0$ vanishes because the populations of the levels $\ket{S,1/2}$ and $\ket{S,-1/2}$ are identical. 
We therefore obtain an interference-based quantum synchronization blockade for $\gamma_\mathrm{g} = \gamma_\mathrm{d}$ for half-integer spins, too. 
Unlike in the case of integer spin, however, the structure of the limit cycle changes depending on the ratio of the dissipation rates [an example for $S=3/2$ is shown in Eq.~\eqref{eqn:LC32}]. 
Consequently, specific values of the imbalance ratio $\gamma_\mathrm{g}/\gamma_\mathrm{d}$ can lead to additional revivals of synchronization blockade when the specific values of the population differences accidentally compensate the amplitude and phase differences due to the first term in Eq.~\eqref{eqn:CoherenceIntegerSpin}.
For example, for $S=3/2$, the phase distribution to leading order in perturbation theory has roots at $\gamma_\mathrm{g}/\gamma_\mathrm{d} \in \{0.108, 1, 9.25\}$, which reproduces well the numerically observed blockades for $S=3/2$. 
Near the region of the second blockade $\gamma_{\mathrm{d}(\mathrm{g})}/\gamma_{\mathrm{g}(\mathrm{d})} \approx 10^{-1}$, the populations of the states $\ket{3/2,(-)1/2}$ and $\ket{3/2,(-)3/2}$ are maximal. 
While one may in principle expect even more blockades for higher spin, numerically, we only observed up to $3$ blockades for the considered spin numbers $S \leq 3$. 

The discussion of the SymLC stabilization scheme so far leads to two conclusions. 
First, the dissipation-rate dependence of the limit-cycle state for half-integer spin gives rise to additional synchronization blockades at certain imbalance ratios of the dissipation rates. 
They can be removed by a modification of the jump operators such that the limit cycle becomes independent of the dissipation rates even for half-integer $S$, which is discussed in Sec.~\ref{sec:nELC}.
Second, the definition $\varepsilon_1$ of the signal strength causes the semiclassical signal to be overwhelmed by the dissipative limit-cycle stabilization for largely imbalanced dissipation rates. 
This shows that the definition $\varepsilon_1$ (which worked well for the case $S=1$) is too restrictive for $S>1$ and causes the signal to be unnecessarily weak.
In the following section, we remedy this shortcoming by considering a different definition of a weak signal.

\subsection{Different definition of the signal strength}\label{sec:epsilon2}

The analysis of the previous section revealed that the definition $\varepsilon_1$ of a weak signal is too restrictive for higher spins $S > 1$:
In the limit of strongly imbalanced dissipation rates, the signal cannot compete with the dissipative processes suppressing the coherences and the synchronization measure vanishes.
Rather than defining the signal strength only in terms of the dissipation rates entering Eq.~\eqref{eq:me1spin}, it is better to focus on the susceptibility of the limit cycle to deformations caused by an applied signal.
One such measure has been proposed in Ref.~\cite{KoppenhoferPRA2019}, which compares the  rates at which coherences are built up by the signal to their damping rates due to the stabilization of the limit cycle. 
Its definition requires explicit knowledge of the coefficients of the steady-state density matrix expanded in powers of the signal strength $\varepsilon$, i.e., an analytical solution of the steady-state density matrix.

Here, we consider a modified version of this measure which can be evaluated numerically. 
Given a small parameter $\eta \ll 1$ which ensures that the signal is a weak perturbation to the limit-cycle stabilization, the corresponding signal strength $\varepsilon_2$ is implicitly defined by
\begin{align}
    \eta = \frac{||\hat{\rho}(\varepsilon_2)- \hat{\rho}^{(0)}||}{||\hat{\rho}^{(0)}||}~,
    \label{eq:e2def}
\end{align}
where $\hat{\rho}^{(0)} \equiv \hat{\rho}(\varepsilon = 0)$ is the limit-cycle state, and $||\hat{O}|| = \sqrt{\tr(\hat{O}^\dagger \hat{O})}$ is the Hilbert-Schmidt norm.
The right-hand side of Eq.~\eqref{eq:e2def} can be evaluated numerically for different signal strengths $\varepsilon$ and the value $\varepsilon_2$ satisfying Eq.~\eqref{eq:e2def} can be found by interpolation.  
In the following, we choose $\eta = 0.01$.

Using this definition of the drive strength $\varepsilon_2$, we show in Fig.~\ref{fig:ELC_epsilon1} (red dashed line) the resulting phase distribution for $S=3/2$. 
The blockade regimes remain the same but the suppression of $S(\phi)$ at largely imbalanced dissipation rates has been lifted. 
Compared to $\varepsilon_1$, the maximum of $S(\phi)$ is larger with the $\varepsilon_2$ definition, and comparable to that of $S=1$ at large dissipation rates.

\section{Gain-loss-asymmetric limit-cycle stabilization}\label{sec:nELC}
\label{sec:GainLossAsymmetricLC}

In Sec.~\ref{sec:ELC}, we found that a semiclassical drive combined with the SymLC stabilization scheme leads to qualitatively different synchronization blockade patterns depending whether the spin $S$ is integer or half-integer. 
For integer spins, the limit cycle is always the $\ket{S,0}$ state and a single synchronization blockade occurs at $\gamma_\mathrm{g} = \gamma_\mathrm{d}$.
For half-integer spins, the structure of the limit cycle depends on the ratio $\gamma_\mathrm{g}/\gamma_\mathrm{d}$ such that two additional blockades at specific ratios of the dissipation rates occur. The additional blockades are due to the fact that the structure of the limit cycle depends on the dissipation rates.

This dependence of the limit cycle on the dissipation rates for half-integer spin can be eliminated by considering a generalized set of jump operators defined as $\hat{O}_\mathrm{g}^M= \hat{S}_{+}(\hat{S}_z - M)$ and $\hat{O}_\mathrm{d}^M= \hat{S}_{-}(\hat{S}_z - M)$.
Now, the jump operators $\hat{O}_\mathrm{g}^M$ ($\hat{O}_\mathrm{d}^M$) do not provide any transitions up (down) from the level $\ket{S,M}$, as shown in Fig.~\ref{fig:spin32Lvls}(c) for $M=-1/2$.
Similar to the SymLC scheme for integer spin, we thus have a unique limit cycle $\ket{S,M}$, which is independent of the ratio $\gamma_\mathrm{g}/\gamma_\mathrm{d}$.
A nonzero value of $M \neq 0$ breaks the symmetry between dissipation rates going up and down the ladder of spin states, which we identified previously in the SymLC scheme.
Therefore, we call this scheme the \emph{gain-loss-asymmetric limit-cycle} (AsymLC) stabilization scheme.
The dissipators used previously for the SymLC scheme correspond to the case $M=0$, i.e., $\hat{O}_\mathrm{g,d} = \hat{O}_\mathrm{g,d}^{M=0}$.

\begin{figure}[t]
	\centering
    \includegraphics[width=\columnwidth]{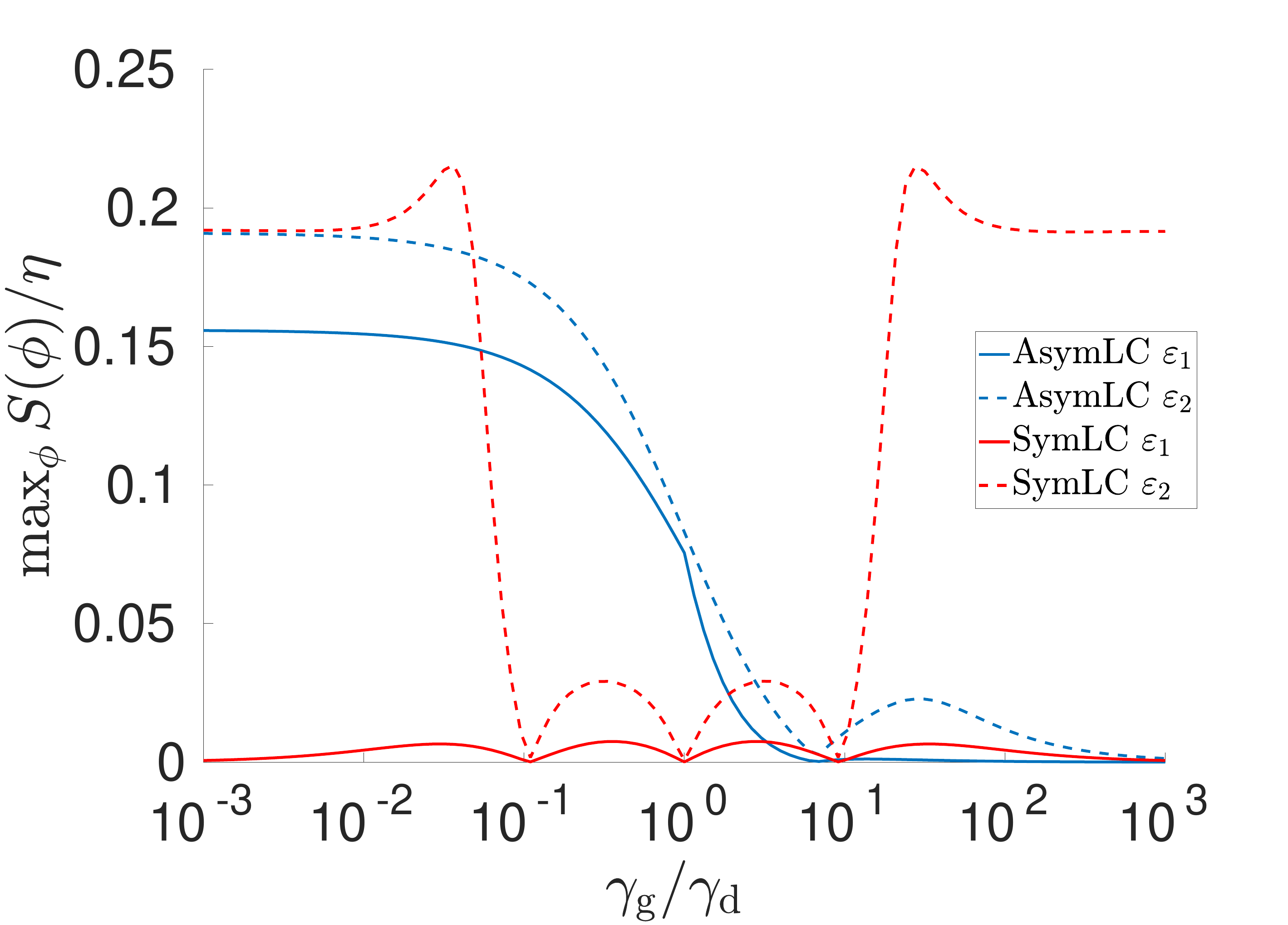}
	\caption{Maximum of phase distribution $S(\phi)$ vs. ratio of dissipation rates $\gamma_\mathrm{g}/\gamma_\mathrm{d}$ for spin $S=3/2$. The red lines show the case of the gain-loss symmetric LC stabilization (SymLC), while the blue lines show the case of stabilization to the $\ket{m=-1/2}$ state (AsymLC).
	An external semiclassical signal is applied and the other parameters are $\eta = 0.01$ and $\Delta = 0$.}
	\label{fig:spin32_LC_compare}
\end{figure}

One may expect that the synchronization blockade of this AsymLC in a half-integer spin system should follow the same pattern as the blockade of a SymLC in an integer spin system.   
Interestingly, this is not the case: 
In Fig.~\ref{fig:spin32_LC_compare}, we compare the maximum of $S(\phi)$ over a large range of dissipation rates between different limit cycles for $S=3/2$. 
At $\gamma_\mathrm{g} = \gamma_\mathrm{d}$, the blockade has been lifted for both $\varepsilon_1$ and $\varepsilon_2$, with the synchronization measure increased by two orders of magnitude 
compared to its small residual value for a SymLC (which is due to higher-order effects in the signal strength).
For $\gamma_\mathrm{g} \ll \gamma_\mathrm{d}$, the maximum of $S(\phi)$ approaches an asymptotic value, whereas in the opposite limit, the synchronization measure is strongly suppressed. 
Around $\gamma_\mathrm{g}/\gamma_\mathrm{d} \approx 7$, a single synchronization blockade remains.

This single synchronization blockade at $\gamma_\mathrm{g}/\gamma_\mathrm{d} \approx 7$ is the equivalent of the unique synchronization blockade at $\gamma_\mathrm{g} = \gamma_\mathrm{d}$ for integer spins, which can be seen as follows. 
The relation ~\eqref{eqn:CoherenceIntegerSpin} can be generalized to the case of the new dissipators $\hat{O}_\mathrm{g,d}^M$ (an explicit expression is given in App.~\ref{app:A}). 
The appearance of the constant $M$ in the recursion relation breaks the inversion symmetry which we identified in the case of integer spin. 
As a consequence, no interference-based synchronization blockade is expected for balanced dissipation rates $\gamma_\mathrm{g} = \gamma_\mathrm{d}$. 
However, we still expect a synchronization blockade for some ratio $\gamma_\mathrm{g}/\gamma_\mathrm{d}$ since we essentially reproduced the limit-cycle physics of an integer spin, but now shifted and centered around the level $\ket{S,M}$:
The limit cycle is independent of the dissipation rates and only the population $\rho_{M,M} = 1$ is nonzero such that, to leading order in the signal strength $\varepsilon$, only the coherences between the level $\ket{S,M}$ and $\ket{S,M \pm 1}$ are built up,
\begin{align}
	\rho_{M+1,M} &= - \frac{\varepsilon A^+_M}{2 i \Delta + \gamma_\mathrm{d} (A^+_M)^2 + \gamma_\mathrm{g} (A^+_{M+1})^2} ~, \\
	\rho_{M,M-1} &= + \frac{\varepsilon A^+_{M-1}}{2 i \Delta + \gamma_\mathrm{d} (A^+_{M-2})^2 + \gamma_\mathrm{g} (A^+_{M-1})^2} ~.
\end{align}
Due to the broken inversion symmetry, we now have to take into account the numerical weights $d^S_{m,m'}$ of the coherences in the synchronization measure $S(\phi)$, i.e, the condition for an interference-based synchronization blockade is
\begin{align}
	S(\phi) = 0 \Leftrightarrow d^S_{M,M+1} \rho_{M+1,M} + d^S_{M-1,M} \rho_{M,M-1} = 0~,
\end{align}
where the numerical weights are defined in App.~\ref{app:A}. 
For $\Delta=0$, $S=3/2$, and $M=-1/2$, the unique solution of this condition is $\gamma_\mathrm{g}/\gamma_\mathrm{d} = 20/3 \approx 6.67$, which matches the observed blockade in Fig.~\ref{fig:spin32_LC_compare}. 

Another feature in Fig.~\ref{fig:spin32_LC_compare} is the very different level of synchronization for largely imbalanced dissipation rates. 
For $\gamma_\mathrm{g}/\gamma_\mathrm{d} \ll 1$, we observe a high level of synchronization, similar to the case of a SymLC and $S=1$ in Fig.~\ref{fig:ELC_epsilon1}. 
However, synchronization is suppressed in the opposite limit $\gamma_\mathrm{g}/\gamma_\mathrm{d} \gg 1$, similar to the case of a SymLC and $S=2$.
For $M=-1/2$, this 
feature is unique to the case of $S=3/2$ and stems from the fact that the limit-cycle state $\ket{3/2,-1/2}$ is only one level away from the bottom end of the spin ladder. 
Population in the level $\ket{3/2,-3/2}$ can only decay via a gain process [see Fig.~\ref{fig:spin32Lvls}(b)], therefore, the coherence $\rho_{-1/2,-3/2}$ depends only on $\gamma_\mathrm{g}$:
\begin{align}
	\rho_{-1/2,-3/2} &= + \varepsilon \frac{2 \sqrt{3}}{4 i \Delta + 3 \gamma_\mathrm{g}}~.
	\label{eqn:AsymLCOneAboveBottom32}
\end{align}
This dependence of the population on only a single dissipation rate is similar to the case of a SymLC in a $S=1$ system. 
In contrast, population in the level $\ket{3/2,1/2}$ decays both via gain and dissipative processes, therefore, the coherence $\rho_{1/2,-1/2}$ depends on both dissipation rates, 
\begin{align}
	\rho_{1/2,-1/2} &= - \varepsilon \frac{4}{4 i \Delta + 4 \gamma_\mathrm{d} + 3 \gamma_\mathrm{g}}~,
\end{align}
like coherences in a SymLC stabilization scheme for $S = 2$. 
Focusing for simplicity on the $\varepsilon_1$ definition of the signal strength, the coherences on resonance thus scale like
\begin{align}
	\rho_{-1/2,-3/2} &\propto \frac{\min(\gamma_\mathrm{g},\gamma_\mathrm{d})}{\gamma_\mathrm{g}} ~, \\
	\rho_{1/2,-1/2} &\propto \frac{\min(\gamma_\mathrm{g},\gamma_\mathrm{d})}{\max(\gamma_\mathrm{g},\gamma_\mathrm{d})}~.
\end{align}
As a consequence, $\rho_{1/2,-1/2}$ is always suppressed for largely imbalanced dissipation rates, but $\rho_{-1/2,-3/2}$ tends to a constant value if $\gamma_\mathrm{g} \ll \gamma_\mathrm{d}$ and $\max_\phi S(\phi)$ remains finite. 
In contrast, for higher half-integer spins $S \geq 5/2$ stabilized to the same state $\ket{S,-1/2}$, synchronization will decrease in the limit $\gamma_\mathrm{g}/\gamma_\mathrm{d} \ll 1$, too.

\begin{figure}[t]
	\centering
    \includegraphics[width=\columnwidth]{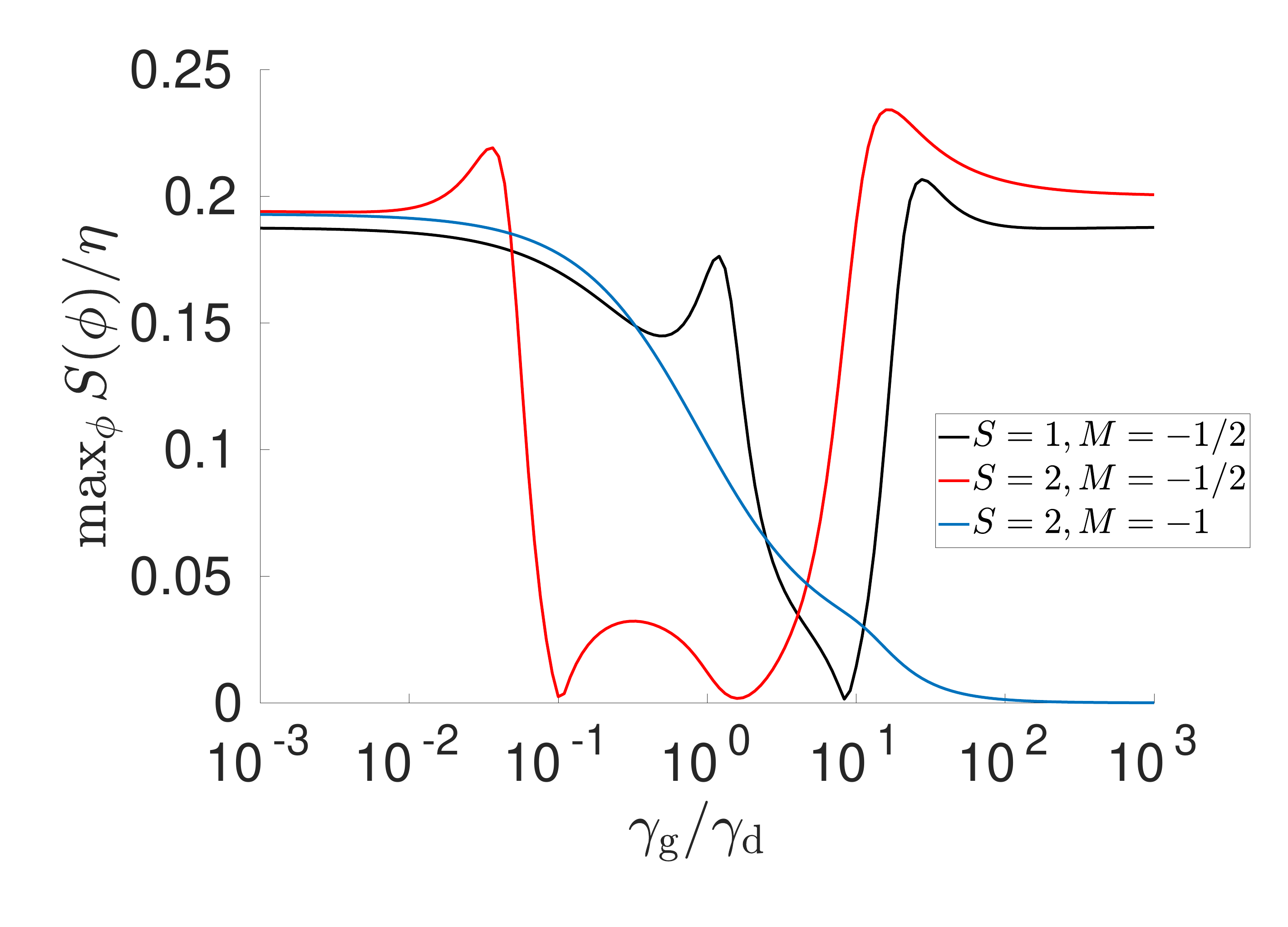}
	\caption{Maximum of phase distribution $S(\phi)$ vs. ratio of dissipation rates $\gamma_\mathrm{g}/\gamma_\mathrm{d}$ for integer spins $S=1$ (black) and $S=2$ (red) stabilized by the AsymLC scheme for $M=-1/2$, and $S=2$ stabilized stabilized by the AsymLC scheme for $M=-1$, i.e., to the state $\ket{m=-1}$ (blue). All systems are driven by an external semiclassical signal with strength $\varepsilon_2$. 
	The other parameters are $\eta = 0.01$ and $\Delta = 0$.}
	\label{fig:IntegerSpinM12}
\end{figure}

It is, likewise, possible for integer spins to be stabilized differently by setting a nonzero value of $M$, resulting in an AsymLC stabilization scheme. The modified jump operators $\hat{O}^{-M}_{g/d}= \hat{S}_{\pm}(\hat{S}_z+M)$ stabilize an integer-spin system to the state $\ket{S,-M}$ for integer-valued $M$; while stabilizing a dissipation-rate-dependent limit cycle for half-integer-valued $M$. We show instances of the AsymLC stabilization scheme for integer spins $S=1$ and $2$ in Fig.~\ref{fig:IntegerSpinM12}. For $S=2$ stabilized to the state $\ket{S,-1}$, we observe large synchronization for $\gamma_\mathrm{g}/\gamma_\mathrm{d} \ll 1$ and suppressed synchronization in the opposite limit $\gamma_\mathrm{g}/\gamma_\mathrm{d} \gg 1$, which is expected along similar reasoning as explained in the case of $S=3/2$. For the case of the stabilization scheme $\hat{O}_{g/d}^{M=-1/2}$, the limit cycles for $S=1$ and $S=2$ are dependent on the ratio of the dissipation rates. The $S=2$ system retains some qualitative features of a SymLC-stabilized $S=3/2$ system: two regions of synchronization blockade similar to the second blockade in the case $S=3/2$, and large asymptotic synchronization values at opposite extremes of the dissipation ratios. For $S=1$, we observe a single point of blockade similar to that in the case of SymLC at balanced rates. 
In both cases, the nonzero value of $M$ breaks the inversion symmetry, resulting in the skewed nature of the synchronization measure along the dissipation-ratio axis. 

\section{Comparison of different limit cycles}
\label{sec:Compare}
\begin{figure}[tb!]
	\centering
    \includegraphics[width=\columnwidth]{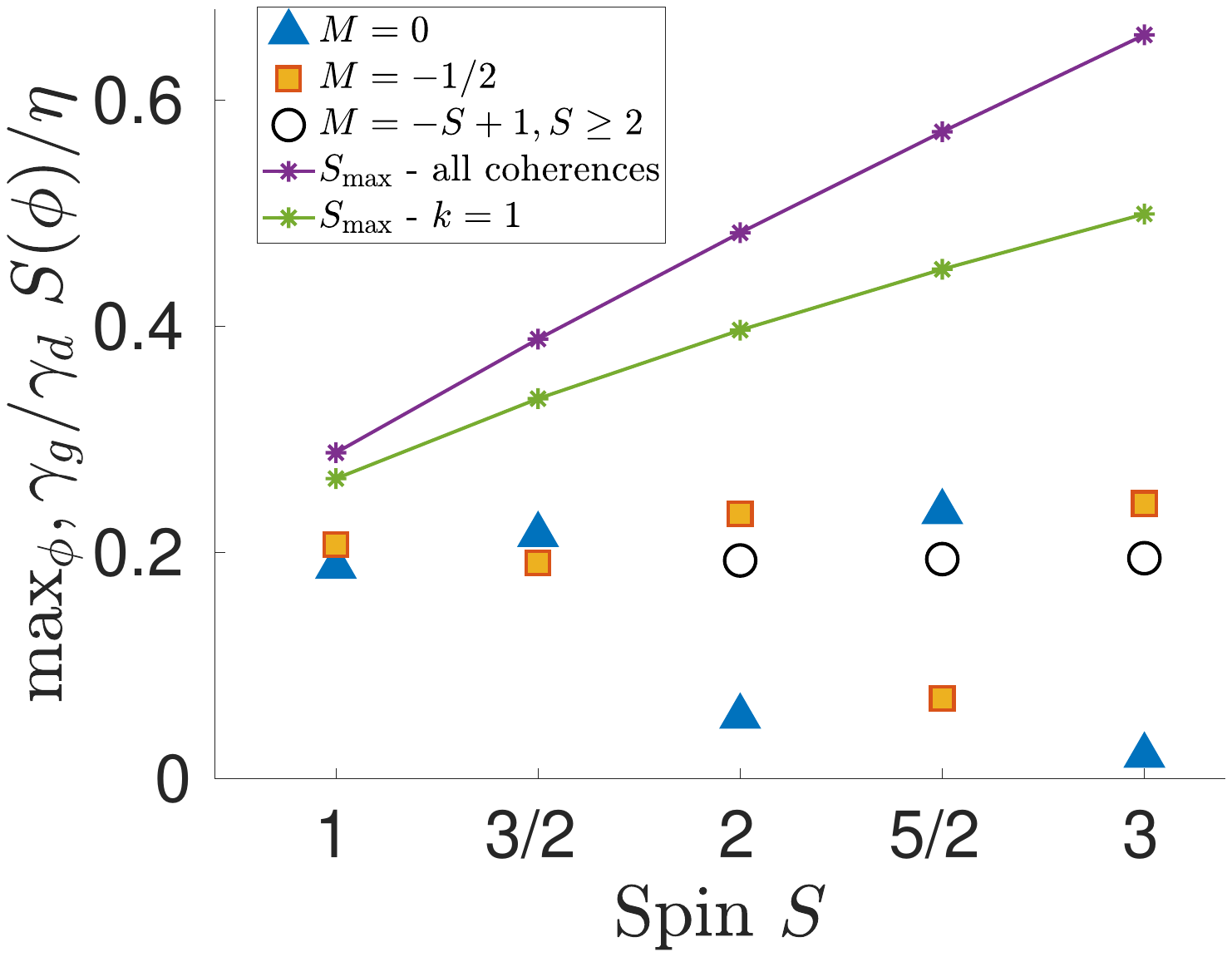}
	\caption{
	Maximum of $S(\phi)/\eta$ for different spin values $S$ and different limit cycles defined by the set of operators $\hat{O}_\mathrm{g}^M= \hat{S}_{+}(\hat{S}_z - M)$ and $\hat{O}_\mathrm{d}^M= \hat{S}_{-}(\hat{S}_z - M)$ for $M=0$ (blue triangles), $-1/2$ (orange squares) and $-S+1$ (black empty circles), optimized over a large range of dissipation ratios $\gamma_\mathrm{g}/\gamma_\mathrm{d}$. For integer (half-integer) spins, integer (half-integer) $M$ results in a dissipation-rate-independent limit-cycle state $\ket{S,M}$; whereas integer (half-integer) spins stabilized with half-integer (integer) $M$ result in a dissipation-rate-dependent limit cycle. In contrast, stabilizing to the state $\ket{S,-S+1}$ results in almost the same value in the measure of synchronization for all values of $S>1$ studied here. 
	The upper bound of $\max_{\phi,\gamma_{g}/\gamma_\mathrm{d}} S(\phi)/\eta$ given by Eq.~\eqref{eqn:Bound} is represented by purple asterisk symbols and takes into account all coherences of the spin-$S$ system. 
    By taking into account only coherences that contribute to the $\cos(\phi)$ term in $S(\phi)$ (which are the only coherences a semiclassical signal can build up to leading order in $\varepsilon$), this bound can be lowered, which is represented by the green asterisk symbols.
	All limit-cycle oscillators are driven by a semiclassical signal and the other parameters are $\eta = 0.01$ and $\Delta = 0$.}
	\label{fig:maxSphivsS}
\end{figure}

We summarize the results on the maximum of $S(\phi)$ for the different oscillators in Fig.~\ref{fig:maxSphivsS}. Here, we use $\varepsilon_2$ as the definition of the drive strength, and we optimize both over $\phi$ and over a large range of dissipation ratios $\gamma_\mathrm{g}/\gamma_\mathrm{d}$, i.e., for each limit-cycle oscillator we choose the ratio $\gamma_\mathrm{g}/\gamma_\mathrm{d}$ that maximizes 
$\max_\phi S(\phi)$. 
For a fixed limit-cycle stabilization scheme (i.e., fixed $M$ in the jump operators $\hat{O}_{g/d}^{M}$), integer $M$ gives rise to a dissipation-rate-independent limit cycle for integer spins and a dissipation-rate dependent limit cycle for half-integer spins, and vice versa for half-integer $M$. 
These two types of limit cycle achieve very different maximum levels of synchronization for integer vs.\ half-integer spin number,
as shown by the large oscillations in the data points of the same type in Fig.~\ref{fig:maxSphivsS}.

In general, a dissipation-rate-dependent limit cycle gives a larger synchronization measure than a dissipation-rate-independent limit cycle for all spins considered in this plot, for $M=0$ (blue triangles) and $M=-1/2$  (orange squares).
Intuitively, this can be understood by considering how many coherences can contribute to the synchronization measure. 
For a limit cycle given by a single energy eigenstate $\ket{M}$ (e.g., generated by the SymLC scheme for integer $S$ or the $M=-1/2$ AsymLC scheme for half-integer $S$), only the two coherences $\rho_{M,M\pm 1}$ can in principle contribute to the synchronization measure. 
Moreover, as discussed in Sec.~\ref{sec:CoherenceGeneration}, they have opposite phases such that one of them has to be suppressed by choosing largely imbalanced dissipation rates. 
In contrast, for a limit cycle that is a mixture of different energy eigenstates (e.g., generated by the $M=-1$ AsymLC scheme for integer $S$ or the SymLC scheme for half-integer $S$), a semiclassical signal can in principle build up all coherences on the first upper and lower diagonals.
By optimizing the ratio $\gamma_\mathrm{g}/\gamma_\mathrm{d}$ to maximize their constructive interference, one can thus achieve a higher level synchronization.

It is interesting to note that when one considers a limit cycle stabilized to the state $\ket{m=-S+1}$ (black empty circles), one reaches a comparable level of synchronization as in the case of the dissipation-rate-dependent limit cycles 
for all investigated spin numbers $S$, even though dissipation-rate-independent limit cycles show very low synchronization for $S \geq 2$. 
This is a generalization of the boundary effect discussed in Sec.~\ref{sec:GainLossAsymmetricLC}:
If the limit-cycle state is only one step away from the bottom (top) of the spin ladder, the coherence between this state and the state at the bottom (top) of the spin ladder depends only on a single dissipation rate $\gamma_\mathrm{g}$ ($\gamma_\mathrm{d}$) [as shown in Eq.~\eqref{eqn:AsymLCOneAboveBottom32} for $S=3/2$ and $M=-1/2$], and this coherence dominates the synchronization behavior. 

Figure~\ref{fig:maxSphivsS} shows that specific combinations of limit cycle and signal exhibit a pronounced integer vs.\ half-integer effect.
However, 
if one considers the maximum level of synchronization across all limit-cycle stabilization schemes for a given spin number $S$,
all spin systems considered here are able to synchronize equally well. This leads to the conjecture that a limit-cycle-agnostic and signal-agnostic optimization may show no difference in synchronization behavior between integer and half-integer spins $S$.
To further support this conjecture and to put the numerical values of the optimized synchronization measure shown in Fig.~\ref{fig:maxSphivsS} into perspective, 
we derive in App.~\ref{app:C} an upper bound on the synchronization measure 
by optimizing over $\phi$ and all possible combinations of limit cycle and signal in a spin-$S$ system,
\begin{align}
	S_\mathrm{max} 
	&= \max_{\mathrm{LC},\mathrm{signal},\phi} S(\phi) 
	= \sqrt{2} \eta \sqrt{\sum_{k=1}^{2S} \sum_{m=-S}^{S-k} (d^S_{m,m+k})^2} ~,
	\label{eqn:Bound}
\end{align}
where $d^S_{m,m'}$ are the coefficients of the synchronization measure defined in App.~\ref{app:A}. 
This bound assumes that all coherences 
that contribute to the same $\cos(k \phi)$ term in $S(\phi)$ interfere constructively and that the relative strength of the coherences contributing to terms with different $k$ are chosen optimally.
The corresponding results are shown by the purple asterisk symbols and line in Fig.~\ref{fig:maxSphivsS}. 
We also plot a variant of the bound where only the $k=1$ term in Eq.~\eqref{eqn:Bound} is kept, i.e., only coherences $\rho_{m,m\pm 1}$ on the first off-diagonal are nonzero, like for a semiclassical signal.
The corresponding data is shown by the green asterisk symbols and line. 
Note that $S_\mathrm{max}$ does not show any integer vs.\ half-integer oscillations.

\section{Conclusion}\label{sec:conclusions} 

In conclusion, we investigated quantum synchronization of a single spin $S\geq 1$ driven by an external semiclassical signal.
Using numerical simulations for spin values up to $S=3$, we showed that half-integer spins synchronize differently under the SymLC stabilization scheme.
More specifically, half-integer spins show additional quantum interference-based synchronization blockades at certain imbalanced ratios $\gamma_\mathrm{g}/\gamma_\mathrm{d}$ of the dissipation rates. 
We presented an analytical description of this phenomenon, which reveals that a synchronization blockade requires the interplay between (i) the generation of coherences by the signal and (ii) the distribution of populations in the limit cycle state. 
If only one of these factors changes as a function of the dissipation rates, only a single synchronization blockade is observed. 
For a SymLC stabilization scheme, it occurs at balanced dissipation rates, $\gamma_\mathrm{g} = \gamma_\mathrm{d}$, but its position can be moved by using an AsymLC stabilization scheme.

We also showed that a simple definition of the signal strength based on the minimum of the dissipation rates is too restrictive for spin systems with higher $S \geq 1$. 
To achieve higher levels of quantum synchronization, we proposed a more refined definition of the signal strength which quantifies the actual deformation of the limit cycle in the presence of the signal. 

More fundamentally, our results highlight that one should naturally consider different stabilization schemes for different spin numbers to achieve better quantum synchronization. 
While the SymLC and AsymLC stabilization schemes show large variations in the maximum synchronization achievable for all $S$ between 1 and 3 and pronounced oscillations between integer and half-integer spin number $S$, 
an appropriate choice of the stabilization scheme for each spin $S$ 
leads to comparable values of the synchronization measure. 
This raises the interesting conjecture that the maximum amount of synchronization is a monotonic function of the spin number if one has full control over the limit cycle (i.e., one can realize dissipators that will stabilize an arbitrary target state), as indicated by an upper bound on the synchronization measure. 
Progress in this direction could be made by proving the tightness of the bound on the synchronization measure obtained by an optimization over all possible limit-cycles and signals, which would also shed light on the quantum-to-classical transition in synchronization for very large spin number $S$.

\begin{acknowledgments}
R.T. and C.B. acknowledge financial support by the NCCR
QSIT, a National Centre of Excellence in Research, funded
by the Swiss National Science Foundation (grant number
51NF40-185902) and by the Swiss National Science Foundation through grant number 200020\_200481. 
\end{acknowledgments}


\appendix

\section{Synchronization measure}
\label{app:A}

In this appendix, we derive the analytical form of the marginal phase distribution $S(\phi)$, as defined in Eq.~\eqref{eq:Sq1},
\begin{align}
    S(\phi) &= \int^\pi_0 d\theta \sin\theta\: Q(\theta,\phi) -\frac{1}{2\pi}~, 
\end{align}
where the Husimi $Q$ function is defined in Eq.~\eqref{eq:Qfunc}, 
\begin{align}
	Q(\theta,\varphi) = \frac{2S+1}{4\pi} \bra{\theta,\phi}\hat{\rho} \ket{\theta,\phi}~.
\end{align}
The spin-coherent states are given by \cite{KoppenhoferPRA2019,BrinkSatchler1968}
\begin{align}
	\ket{\theta,\phi} 
	&= e^{-i \phi \hat{S}_z} e^{-i \theta \hat{S}_y} \ket{S,S} \nonumber \\
	&= \sum_{m=-S}^S D^S_{m,S}(\phi,\theta,0) \ket{S,m} ~,
\end{align}
where we defined
\begin{align}
	D^S_{m,S}(\alpha,\beta,\gamma) = \bra{S,m} e^{-i \alpha \hat{S}_z} e^{-i \beta \hat{S}_y} e^{-i \gamma \hat{S}_z} \ket{S,S} ~.
\end{align}
The elements of the Wigner $D$ matrix can be rewritten as $D^S_{m,S}(\phi,\theta,0) = e^{-i m \phi } D^S_{m,S}(0,\theta,0)$, which can be further evaluated using the relation \cite{Wigner1959}
\begin{align}
	&D^S_{m,S}(0, \theta, 0) \nonumber \\
	&= \sqrt{\frac{(2S)!}{	(S+m)!(S-m)!}} \cos\left( \frac{\theta}{2} \right)^{S+m} \sin \left( \frac{\theta}{2} \right)^{S-m} ~.
\end{align}
The explicit expression for the Husimi $Q$ function is thus
\begin{align}
	Q(\theta,\varphi) &= \sum_{m,m'} e^{-i  (m-m')\phi} q^S_{m,m'} \rho_{m',m} ~, \\
	q^S_{m,m'} &= \frac{2S+1}{4 \pi}  D^S_{m,S}(0,\theta,0) D^S_{m',S}(0,\theta,0)~.
\end{align}
Using this result, we now perform the integration in $S(\phi)$, 
\begin{align}
	\int_0^\pi \mathrm{d} \theta\, \sin \theta\: Q(\theta,\phi) 
    &= \sum_{m,m'} e^{-i (m - m')\phi} c^S_{m,m'} \rho_{m',m}~,
\end{align}
where the coefficients $c^S_{m,m'}$ are given by
\begin{align}
	&c^S_{m,m'} 
	= \int_0^\pi \mathrm{d} \theta\, \sin \theta\: q_{m,m'}^S 
	= \frac{2S+1}{4 \pi} \nonumber \\
	&\times \frac{2 (2S)! \Gamma \left( 1 + S - \frac{m + m'}{2} \right) \Gamma \left(1 + S + \frac{m + m'}{2} \right)}{\sqrt{(S-m)!(S+m)!(S-m')!(S+m')!} \Gamma(2 + 2S)}~,
\end{align}
where $\Gamma(x)$ is the Gamma function. 
Note that $c^S_{m,m} = 1/2 \pi$.
Using the definition Eq.~\eqref{eq:Sq1}, we therefore find
\begin{align}
	S(\phi) &= \sum_{m,m'} e^{-i (m-m')\phi} d^S_{m,m'} \rho_{m',m} ~, \displaybreak[1]\\
	d^S_{m,m'} &= c^S_{m,m'} (1 - \delta_{m,m'}) ~.
	\label{eqn:Weights}
\end{align}

\section{Density-matrix elements for the SymLC and AsymLC stabilization scheme}
\label{app:B}

In this appendix, we derive the density matrix elements of the steady state of the Lindblad master equation:
\begin{align}
	\frac{\mathrm{d}}{\mathrm{d} t} \hat{\rho} = - i [\hat{H},\hat{\rho}] + \gamma_\mathrm{g} \mathcal{D}[\hat{O}_\mathrm{g}^M] \hat{\rho} + \gamma_\mathrm{d} \mathcal{D}[\hat{O}_\mathrm{d}^M] \hat{\rho}~, 
\end{align}
where the Hamiltonian $\hat{H}$ is given by Eq.~\eqref{eq:Ham1spin} and the jump operators defined by $\hat{O}_\mathrm{g}^M= \hat{S}_{+}(\hat{S}_z - M)$ and $\hat{O}_\mathrm{d}^M= \hat{S}_{-}(\hat{S}_z - M)$. These operators stabilize the spin state $\ket{S,M}$, where $M$ takes on integer numbers for integer spins, and half-integer for half-integer spins. When $M=0$, we recover the SymLC stabilization scheme given by Eq.~\eqref{eq:me1spin}.

Projecting this master equation on the spin eigenstates $\ket{S,m}$, we find the following recursion relation for the steady-state value of a particular coherence $\rho_{n,m}$:
\begin{widetext}
\begin{align}
	\rho_{n,m} 
	&= \frac{1}{2 i (n - m) \Delta + \gamma_\mathrm{d} \left[ (m-M)^2 (A^-_{m})^2 + (n-M)^2 (A^-_{n})^2 \right] + \gamma_\mathrm{g} \left[ (m-M)^2 (A^+_{m})^2 + (n-M)^2 (A^+_{n})^2 \right]} \nonumber \\
	&\times \Big[ \begin{aligned}[t]
		&2 \gamma_\mathrm{g} (n-1-M) (m - 1 - M) A^-_{m} A^+_{n-1} \rho_{n-1,m-1} + 2 \gamma_\mathrm{d} (n + 1 - M) (m + 1 - M) A^+_{m} A^-_{n+1} \rho_{n+1,m+1} \\
		&+ \varepsilon \left(  A^-_{m+1} \rho_{n,m+1} + A^-_{n+1} \rho_{n+1,m} - A^+_{n-1} \rho_{n-1,m} - A^+_{m-1} \rho_{n,m-1}\right) \Big]~,
	\end{aligned}
	\label{eqn:RecursionRelationCoherencenmM}
\end{align}
where $A^\pm_m = \sqrt{S(S+1) - m (m \pm 1)}$ denotes the matrix elements of the spin raising and lowering operators, $\hat{S}_\pm \ket{S,m} = A^\pm_{m} \ket{S, m \pm 1}$. 
To turn this result into a closed expression for the coherences, we use the fact that synchronization is a perturbative effect in the signal strength $\varepsilon$. 
In the absence of a signal, $\varepsilon = 0$, the limit-cycle state must be a statistical mixture of the spin eigenstates, i.e., all coherences are zero. 
This is shown in Eq.~\eqref{eqn:LC32} for the specific case $S=3/2$. 
The semiclassical signal generates transitions between the $\ket{S,m}$ states, but it can only change the $m$ quantum number in steps of one. 
Therefore, to leading order in the signal strength $\varepsilon$, only the first off-diagonal coherences will be nonzero, $\rho_{n,n\pm 1} = \mathcal{O}(\varepsilon)$, and all other coherences will be suppressed by at least another power of $\varepsilon$, $\rho_{n,n \pm k} = \mathcal{O}(\varepsilon^2)$ for $k \geq 2$. 
In this limit, we thus obtain to first order in $\varepsilon$ the explicit expression
\begin{align}
	\rho_{n,n-1} 
	&= \frac{\varepsilon A^-_n \left( \rho_{n,n} - \rho_{n-1,n-1} \right)}{2 i \Delta + \gamma_\mathrm{d} \left[ (n-M-1)^2 (A^-_{n-1})^2 + (n-M)^2 (A^-_n)^2 \right] + \gamma_\mathrm{g} \left[ (n - M - 1)^2 (A^+_{n-1})^2 + (n - M)^2 (A^+_n)^2 \right]}~,
	\label{eqn:CoherenceFirstOrder}
\end{align}
which generalizes Eq.~\eqref{eqn:CoherenceIntegerSpin} of the main text.
\end{widetext}

\section{Maximum synchronization}
\label{app:C}

In this appendix, we derive an upper bound on the synchronization measure $S(\phi)$ defined in Eq.~\eqref{eq:SphiGeneral} by maximizing over all limit cycles and signals that can be applied in a spin-$S$ system. 
This generalizes a similar discussion given in Ref.~\cite{KoppenhoferPRA2019} to arbitrary spin number $S$.

Our starting point is the following ansatz for the steady-state density matrix in the presence of an external signal, expanded to leading order in the signal strength $\varepsilon$,
\begin{align}
	\hat{\rho}_\mathrm{ss} &= \hat{\rho}_\mathrm{LC} + \varepsilon \hat{\rho}_\mathrm{coh} ~,
	\label{eqn:MaxSync:Ansatz} 
\end{align}
where $\hat{\rho}_\mathrm{LC}$ contains the populations of the density matrix (which are defined by the limit-cycle stabilization scheme), $(\hat{\rho}_\mathrm{LC})_{m,n} = \delta_{m,n} \rho_{m,n}$, and $\hat{\rho}_\mathrm{coh}$ contains the coherences (which are due to the applied signal), $(\hat{\rho}_\mathrm{coh})_{m,n} = (1 - \delta_{m,n}) \rho_{m,n}$.
Using Eq.~\eqref{eq:e2def}, the value of the signal strength $\varepsilon$ ensuring that the signal is a small perturbation to the limit-cycle stabilization dynamics is 
\begin{align}
	\varepsilon 
	= \eta \frac{|| \hat{\rho}_\mathrm{LC} ||}{|| \hat{\rho}_\mathrm{coh}||} 
	= \eta \sqrt{\frac{\sum_{m=-S}^S \rho_{m,m}^2}{2 \sum_{k=1}^{2 S} \sum_{m=-S}^{S-k} \vert\rho_{m+k,m}\vert^2}} ~,
	\label{eqn:MaxSync:SignalStrengthEpsilon}
\end{align}
where $\eta \ll 1$ is the small dimensionless expansion parameter ensuring that higher-order correction terms to Eq.~\eqref{eqn:MaxSync:Ansatz} are negligible.
The synchronization measure~\eqref{eq:SphiGeneral} evaluated for the ansatz~\eqref{eqn:MaxSync:Ansatz} is
\begin{align}
	S(\phi) 
	&= \varepsilon \sum_{k=1}^{2S} \Bigg[ \underbrace{\sum_{m=-S}^{S-k} d^S_{m,m+k} \rho_{m+k,m}}_{a_k} e^{i k \phi} + \mathrm{c.c.} \Bigg] \nonumber \\
	&= 2 \varepsilon \sum_{k=1}^{2S} \vert a_k \vert \cos \left[ k \phi + \arg(a_k) \right] ~, 
	\label{eqn:MaxSync:SyncMeasureEvaluated}
\end{align}
where we defined the weighted sum $a_k$ of all coherences on the $k$-th diagonal above the main diagonal. 
As discussed below Eq.~\eqref{eq:SphiGeneral}, all coherences on the $k$-th diagonal have the same wavenumber $k = m - m'$ and contribute to the same cosine term.

We now combine Eqs.~\eqref{eqn:MaxSync:SignalStrengthEpsilon} and~\eqref{eqn:MaxSync:SyncMeasureEvaluated} and maximize the result over all free parameters, i.e., the phase $\phi$, the distribution of population in the limit cycle, and the magnitude and phase of each coherence, 
\begin{align}
	S_\mathrm{max} 
	= \max_{\{\rho_{m,m'}\}} \Bigg( &\sqrt{2} \eta \sqrt{\sum_{m=-S}^S \rho_{m,m}^2} \nonumber \\
	&\times \frac{\sum_{k=1}^{2S} \vert a_k \vert \cos[ k \phi + \arg(a_k)]}{\sqrt{\sum_{k=1}^{2S} \sum_{m=-S}^{S-k} \vert \rho_{m+k,m}\vert^2}} \Bigg) ~.
\end{align}
Our first step is to adjust the global phase of coherences along the $k$-th diagonal such that each $a_k \geq 0$ is real.
All cosine terms will then have a common maximum at $\phi=0$. 
Next, we adjust the relative phases of coherences along the $k$-th diagonal such that all of them interfere constructively.
We then find
\begin{align}
	&\sqrt{2} \eta \sqrt{\sum_{m=-S}^S \rho_{m,m}^2} \frac{\sum_{k=1}^{2S} \vert a_k \vert \cos[ k \phi + \arg(a_k)]}{\sqrt{\sum_{k=1}^{2S} \sum_{m=-S}^{S-k} \vert \rho_{m+k,m}\vert^2}} \nonumber \\
	\leq &\sqrt{2} \eta \sqrt{\sum_{m=-S}^S \rho_{m,m}^2} \frac{\sum_{k=1}^{2S} \vert a_k \vert}{\sqrt{\sum_{k=1}^{2S} \sum_{m=-S}^{S-k} \vert \rho_{m+k,m}\vert^2}} \nonumber \\
	\leq &\sqrt{2} \eta \sqrt{\sum_{m=-S}^S \rho_{m,m}^2}  \frac{\sum_{k=1}^{2S} \sum_{m=-S}^{S-k} d^S_{m,m+k} \vert \rho_{m+k,m}\vert}{\sqrt{\sum_{k=1}^{2S} \sum_{m=-S}^{S-k} \vert \rho_{m+k,m}\vert^2}} \nonumber \\
	= &\sqrt{2} \eta \sqrt{\sum_{m=-S}^S \rho_{m,m}^2}  \frac{\vec{d} \cdot \vec{\rho}}{\vert\vec{\rho}\vert} ~,
\end{align}
where we defined the vectors $\vec{d}$ and $\vec{\rho}$ containing the weights $\{d^S_{m,m+k}\}$ and the magnitudes $\{ \vert\rho_{m+k,m}\vert \}$ of the coherences.
The last expression is maximized if the magnitudes of the coherences are chosen such that $\vec{d} = \vec{\rho}$, which yields
\begin{align}
	&\sqrt{2} \eta \sqrt{\sum_{m=-S}^S \rho_{m,m}^2}  \frac{\vec{d} \cdot \vec{\rho}}{\vert\vec{\rho}\vert} \nonumber \displaybreak[1]\\
	\leq &\sqrt{2} \eta \sqrt{\sum_{m=-S}^S \rho_{m,m}^2} \sqrt{\sum_{k=1}^{2S} \sum_{m=-S}^{S-k} (d^S_{m,m+k})^2} ~.
	\label{eqn:MaxSync:OptimizationOfCoherencesComplete}
\end{align}
Finally, we optimize over the populations and obtain
\begin{align}
	S_\mathrm{max} = \sqrt{2} \eta \sqrt{\sum_{k=1}^{2S} \sum_{m=-S}^{S-k} (d^S_{m,m+k})^2} ~.
	\label{eqn:MaxSync:UpperBound}
\end{align}
The maximum is achieved if the limit-cycle oscillator is stabilized into a single energy eigenstate, which is the case for the SymLC (AsymLC) stabilization scheme for integer (half-integer) $S$.

Note that, in general, Eq.~\eqref{eqn:MaxSync:UpperBound} only provides an upper bound to the achievable synchronization measure since not all conditions on the optimal values of coherences and populations may be satisfied simultaneously. 
For example, a semiclassical signal can only generate coherences of the form $\rho_{n,n \pm 1}$ to leading order in $\varepsilon$ such that one cannot benefit from the terms with $k \neq 2$. 
Moreover, Eq.~\eqref{eqn:CoherenceFirstOrder} shows that a semiclassical signal can never generate a nonzero coherence $\rho_{n,n-1}$ to leading order in $\varepsilon$ if the corresponding populations are identical, $\rho_{n,n} = \rho_{n-1,n-1}$. 
The optimal population distribution, however, is achieved if the limit-cycle oscillator is in a single energy eigenstate and all other populations are zero.  
We do not discuss the tightness of this bound for arbitrary $S$ here and only note that, for an $S=1$ system, it has been shown that Eq.~\eqref{eqn:MaxSync:UpperBound} is actually a tight upper bound~\cite{KoppenhoferPRA2019}. 
In this case, the optimal spin-$1$ limit cycle is close to an energy eigenstate with a small asymmetry that ensures that all population differences are nonzero.

\bibliography{biblio}
\bibliographystyle{quantum}

\end{document}